\documentstyle[sprocl,epsf]{article}

\newcommand{\eg}{{\it e.g.}}
\def\VEV#1{\left\langle{ #1} \right\rangle}

\begin{document}

\title{T864 (MINIMAX): A SEARCH FOR DISORIENTED CHIRAL CONDENSATE AT
THE FERMILAB COLLIDER}

\author{J. D. BJORKEN\\ (for the MiniMax Collaboration\footnote{See
the MiniMax web pages {\it http://fnmine.fnal.gov} for more information
about the experiment, including a collaboration list,  beautiful
pictures, and links to papers and transparency copies from talks.})}

\address{Stanford Linear Accelerator Center\\
Stanford University, Stanford, CA 94309}

\maketitle
\abstracts{A small test/experiment has been performed at the Fermilab
Collider to measure charged particle and photon multiplicities in the
forward direction, $\eta \approx 4.1$.  The primary goal is to search
for disoriented chiral condensate (DCC).  The experiment and analysis
methods are described, and preliminary results of the DCC search are
presented.}

\section{Introduction}
\label{sec:1}

In this talk I will describe the status of a small test/experiment
(T864 (MiniMax)) designed to search for disoriented chiral condensate
(DCC) and performed over the last three years at the TeVatron collider.
The origins of MiniMax go back earlier to an initiative designed to
provide the SSC with a full-acceptance detector (FAD). \cite{ref1}
During the associated workshop activity, it was acutely realized that
some of the physics goals were accessible already at Fermilab.  A
collaboration was created (MAX) and their proposal was considered in
the fall of 1992 by the Fermilab program committee, but was rejected.
We decided not to give up, reduced the scope considerably, and on April
1, 1993, resubmitted the revised MiniMax proposal. It was conditionally
approved by the director in late May of that year.

The experimental goals of MiniMax are as follows: 
\begin{enumerate}
\item
Demonstrate that experimentation in the far-forward direction in
collider mode is feasible.
\item
Search for the  anomalies (Centauro, anti-Centauro (JACEE)) 
reported by the cosmic-ray community in this region of phase space.
\item
Search for disoriented chiral condensate.
\item
Contribute to general multiparticle-production phenomenology.
\end{enumerate} 

\section{Physics}
\label{sec:2}
  
The primary motivation for our enterprise has been the search for DCC.
By DCC we mean a piece of strong-interaction vacuum with a rotated
value of its chiral order parameter. The QCD vacuum contains a boson
condensate, like the Higgs sector of electroweak theory. This
condensate arises as a consequence of the spontaneous breaking of the
chiral $SU(2)\times SU(2) = O(4)$ symmetry of QCD. The collective
excitations of this condensate are the pions (Goldstone bosons), and
would be strictly massless were the chiral symmetry exact. The
condensate transforms as a 4-vector ($\sigma, \vec\pi$) and in ordinary
vacuum points in the sigma direction. But perhaps in the interior of
high-energy collision fireballs the orientation is different. If so,
the piece of disoriented vacuum will eventually decay into true vacuum,
and the decay products will be a pulse of coherent semiclassical pion
field carrying the quantum numbers of the disoriented vacuum. In
particular all decay pions in a given event will have the same
(Cartesian) isospin. This feature leads to the basic signature for DCC
searches, namely large fluctuations in the fraction of produced pions
which are neutral:
\begin{equation}
f = \frac{N_{\pi^0}}{N_{\pi^0}+N_{\pi^+}+ N_{\pi^-}} \ .
\label{eq:a}
\end{equation}
If the orientation of the chiral order parameter is random event-by-event, 
then
\begin{equation}
\frac{dN}{df} = \frac{1}{2\sqrt f} \ ,
\label{eq:b}
\end{equation}
which is very different from the conventional wisdom, as embodied in 
Monte-Carlo event generators. 

The cosmic ray evidence on Centauro and anti-Centauro events 
serves as a motivation for this idea. The cosmic-ray observations 
are sensitive to particles produced in the forward direction at large
cms pseudorapidities. This has also served as motivation for the choice
of acceptance for the MiniMax experiment. The experiment is sensitive
to the same leading-particle region as the JACEE event described by
Lord and Iwai,\cite{ref2} and we have used that event as a prototype of
what might be out there waiting to be observed.

While there has been a great deal of theoretical interest in DCC
production,\cite{ref3} there still is not enough development to create
production models appropriate to our experimental needs. We have done a
small amount of work ourselves, but have been too busy with the
experiment to go very far in that direction. At present the operational
MiniMax definition of DCC is as follows:

DCC is a cluster of pions with near-identical momenta and a
distribution of neutral fraction which follows the DCC
inverse-square-root rule. Consequently, in the DCC rest frame the pions
have low kinetic energy and the mass of the DCC ``snowball" will be
only slightly larger than $N$ times the pion mass, with $N$ the
multiplicity of the pion cluster. A very important parameter is the
mean kinetic energy of these pions in the DCC rest frame. It can be
expected to be quite small, perhaps no larger than the pion rest mass,
provided the proper time of the DCC decay process is relatively large,
leading to a large emission volume or area.

It is very advantageous for us in the MiniMax experiment to search for
DCC produced with high transverse velocity. When this occurs the
products of the DCC ``snowball" are boosted into a ``coreless jet" which
occupies a quite limited region of (lego) phase space. Even this
limited region is at least as large as the MiniMax lego acceptance.
However, for experiments with large acceptance, the question of (lego)
coherence length will enter. It is not clear that, even if DCC is
produced over all of (lego) phase space, the chiral order parameter
will point in a common direction. The characteristic length may be no
larger than one or two lego units, and attention will have to be paid
to this problem in experiments with acceptance large compared to
MiniMax.

In any case, for us the important parameters for the simulation of DCC
are the distribution of multiplicity $N$, the rest-frame kinetic-energy
distribution, the DCC transverse-velocity distribution, and to some
extent the pseudorapidity distribution.

\section{History and Description of the Apparatus}

The location in space of the MiniMax experiment was at the C0 collision
region of the Fermilab TeVatron proton-antiproton collider. The C0
region is precisely halfway between D0 and CDF (B0), the experiments
responsible for the top-quark discovery. The MiniMax location in time
coincided with the top-quark data-acquisition period (1993-1996). From 
proposal submission to removal and dismantling of the apparatus, the time
interval was just under three years. The location of the experiment in
fiscal space was near the imaginary axis.

\begin{figure}[htbp]
\begin{center}
\leavevmode
{\epsfxsize=3.75in\epsfbox{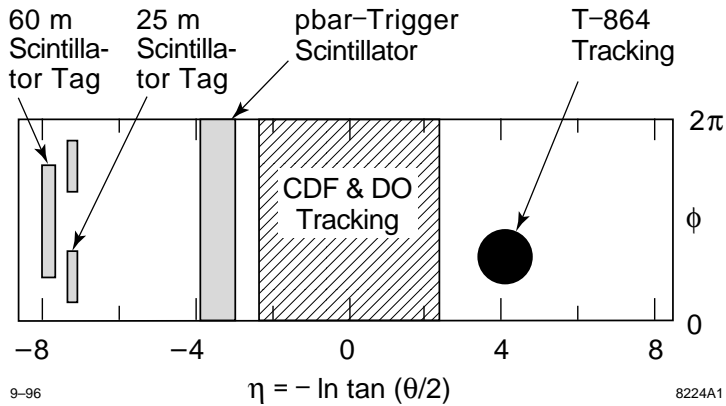}}
\end{center}
\caption{MiniMax lego acceptance.}
\label{fig:b}
\end{figure}

The location in phase space is illustrated in Fig.~\ref{fig:b}. The
heart of the apparatus consisted of a 24-plane MWPC tracking telescope,
backed up with a 28-element lead-scintillator electromagnetic
calorimeter. Behind the eighth MWPC plane was placed a remotely movable
lead converter.  In most runs the lead thickness was chosen to be 1
$X_0$ which allowed photon conversion products to be detected via
tracking information in the rear 16 MWPC planes. Trigger scintillator
was embedded in the telescope as well, at the position of the lead and
just in front of the calorimeter.  The lego acceptance of this
telescope was a circle of radius 0.65 centered at pseudorapidity $\eta
= 4.1$ in the proton hemisphere.

In addition, an 8-element trigger hodoscope was placed around the beam
in the antiproton hemisphere at $\eta \approx -3.0$. Much further
upstream were placed additional counters designed to tag events with
leading antinucleons. Two 10 cm $\times$ 10 cm hadron calorimeter
modules (plus some pieces of scintillator) were placed astride the
beampipe 25 meters upstream in the antiproton direction. One of the
calorimeters was sensitive to 400 $GeV$ antiprotons produced in the
collision and swept into the the calorimeter by accelerator magnets;
the other was sensitive to antineutrons produced at zero degrees.
Despite the poor containment, these detectors provide quite clean tags
for the production of antibaryons of $x_{F} \approx 0.5$. Even
further upstream, at 60 meters in the antiproton direction,
scintillator was placed adjacent to the beampipe. These were sensitive
to $\bar p$'s of $x_{F} \approx 0.85-0.90$ which were swept by the
machine magnets into the beampipe in the proximity of the
scintillators. All these upstream tags were, under most of our running
conditions, very pure. The evidence for this comes from the
multiplicity distribution of the $\eta = -3$ trigger hodoscope, which
shows high sensitivity to the nature of the tag and/or trigger.

In this short report it is not possible to describe in much detail the
actual operating conditions. While it was anticipated that creating a
clean trigger would be difficult, this turned out to be straightforward.
The background rate from beam halo/beam-gas interactions was in most of
our production running lower that 1 percent and well understood. The
trigger cross section was 35--40 $mb$, out of a total nondiffractive,
inelastic proton-antiproton cross section of 50 $mb$. Luminosity was
determined from the D0 luminosity monitors, the known ratio of the
machine beta-functions at C0 and D0, and the known values of the
individual bunch intensities.

The data acquisition system ran at about 50 Hz, and in the production
running in 1995 and early 1996 about 8 million events were
recorded.

The tracking system (about 3000 wires) and its readout performed well,
with wire efficiencies well above 95 percent. The main difficulty is
that the background levels in the chambers, originating from
secondaries from real proton-antiproton collision products, were quite
large. The mean occupancy per wire ranged from 10 percent to 30 percent
depending upon location. The distribution of background hit density
follows a simple rule: directly proportional to the distance from the
luminous region and inversely proportional to the distance from the
beam axis. Despite this large occupancy, we have in the late runs,
comprising over half the data set, been able to reconstruct tracks in
all but the last few percent of the events, thanks to the simple
geometry and the large number of planes.

Simulations of the physics and detector response have been developed.
PYTHIA is used as the event generator for generic events.\cite{ref4} In
addition a DCC event generator has been created, using the definitions
of DCC described in Section~\ref{sec:2}. The particles so created are
tracked via GEANT \cite{ref5} through the detector. The main sources of
background (\eg\ floor, beampipe, apparatus material, etc.) are
included. While the results of the simulation agree well with early
data taken with the original beampipe, later data taken after the
beampipe designed for the experiment was installed in early 1995 do not
agree. The simulation underestimates the observed background by nearly
a factor two; this discrepancy remains not understood.

Track-finding algorithms are still under development. At present, two
different combinatorial trackers are in use. Separate track segments
are constructed for the front eight planes and the back sixteen, with
matching then performed at the position of the lead. As mentioned above
the track-finding efficiency remains high for all but the last few
percent or so of the data set.  There are many candidate algorithms for
finding $\gamma$'s. Our present definition for a gamma conversion
candidate is at least one track originating at the lead which points
toward the luminous region. The simulations are used to estimate the
efficiency of these algorithms. For the $\gamma$'s, the efficiency is
about 65 percent per conversion above a gamma laboratory energy of 3
GeV; below that the efficiency drops off rapidly.

\section{Analysis Strategy}

There are many obstacles for the MiniMax experiment to overcome in
trying to infer the presence or absence of DCC in the data. There is no
momentum information for charged tracks and $\gamma$'s. Not all charged
tracks are pions. Neutral pions are not reconstructed; indeed most of
the time only one of the two $\gamma$'s enter the quite limited MiniMax
acceptance. Only half of the $\gamma$'s convert in the lead.
Efficiencies are of course not 100 percent, and they may be correlated,
in particular, with multiplicity or background level. Nevertheless we
have reason to believe that a meaningful analysis still can be done.

The main reason for our cautious optimism is that we have identified
robust observables which are insensitive to most (but of course not
all) of the aforementioned problems.\cite{ref6} Our raw data consist of
a table of probabilities $P(n_{\rm ch},n_\gamma)$ for finding per event
$n_{\rm ch}$ charged tracks and $n_\gamma$ converted $\gamma$'s. It is,
as is not uncommon in such analyses, convenient to trade in this table
for the table of bivariate normalized factorial moments constructed
from the generating function for $P$:
\begin{equation} 
G(z_{ch},z_\gamma) = \sum z^{n_{ch}}_{ch}\, z^{n_\gamma}_\gamma\,
P\left(n_{ch},n_\gamma\right) \ .
\label{eq:c}
\end{equation} 
We then try to relate this generating function to the ideal one
describing the production of pions, generic or DCC. Most models of pion
production (including our DCC model) assume some parent distribution
$P(N)$ for producing $N$ pions, followed by a binomial distribution for
the partition into charged and neutral pions. We call this hypothesis
generic pion production.
\begin{equation} 
p(n_{ch},n_0) = P(N){N\choose n_0}\, f^{n_0}
(1-f)^{n_{ch}} \qquad N = n_{ch}+ n_0\qquad f \approx \frac{1}{3} \ .
\label{eq:d}
\end{equation} 
From these equations, it is easy to work out that the bivariate
generating function describing production of charged and neutral pions
is obtained from the single-variable generating function for the parent
pions by replacing the pion fugacity variable $z$ by a linear
combination of the fugacities for the charged and neutral pions,
weighted by the assumed neutral fraction $f \cong 1/3$. That is, if
\begin{equation}
G(z) = \sum z^NP(N)
\label{eq:e}
\end{equation}
then
\begin{equation} 
G(z_{ch},z_0) = G(f z_0+(1-f)\, z_{ch})\ .
\label{eq:f}
\end{equation} 
Furthermore if the efficiency $\epsilon_{\rm ch}$ for finding a charged
track from a parent charged pion and the efficiencies $\epsilon_0$,
$\epsilon_1$, $\epsilon_2$ for our finding 0, 1, or 2 $\gamma$'s
respectively are uncorrelated with multiplicity or environment, then the
generating function for the MiniMax observables as defined above will be
again obtained by replacing the pion fugacities by weighted
fugacities: \cite{ref7}
\begin{eqnarray} 
z_{ch} &\rightarrow& \epsilon\, z_{ch} + (1-\epsilon) \nonumber \\[1ex]
z_0 &\rightarrow& \epsilon_0 + \epsilon_1 z_\gamma + \epsilon_2
z^2_\gamma \ .
\label{eq:g}
\end{eqnarray} 
The main consequence of these convolutions is that, within the above
assumptions, the generating function for the MiniMax bivariate moments
is actually only a function of a single variable, not two, if the
underlying production dynamics is generic. Therefore there must be many
relations between the elements of the array of bivariate factorial
moments measured by MiniMax, if generic production prevails. And it
turns out that many of these relations are independent of the
efficiency factors introduced above. Especially robust variables turn
out to be ratios of the normalized bivariate factorial moments. For
example, the quantities
\begin{equation} 
R_{ij} = \frac{F_{ij}}{F_{i+j,0}}
\label{eq:h}
\end{equation} 
with
\begin{equation}
F_{ij} = \frac{1}{\VEV{n_{ch}}^i\VEV{n_\gamma}^j}\ 
\left.\left(\frac{\partial}{\partial z_{ch}}\right)^i 
\left(\frac{\partial}{\partial z_\gamma}\right)^jG(z_{ch},z_{\gamma})\right|
_{z_{ch}= z_\gamma=1}\label{eq:i}
\end{equation}
can all be shown to equal unity when $j=1$, if efficiencies are
uncorrelated and the pion production is generic. In general, the
remaining $R_{ij}$ depend only on one additional parameter,
proportional to the probability that both $\gamma$'s from a parent
$\pi^0$ are detected.

What happens if the production mechanism is not generic, but DCC? Then
one can work out, not quite as easily as for generic production,
Eq.~(4), that all one has to do is to take the binomial distribution
with neutral fraction $f$ and integrate $f$ over the
inverse-square-root weight to get the DCC distribution. This leads to a
MiniMax bivariate generating function which depends nontrivially on two
variables. But again when calculating the above ratios $R_{ij}$ the
uncorrelated efficiency factors do not appear, and the values of the
$R_{ij}$ are nowhere near unity. For example,
\begin{equation}
R_{i1} = \frac{1}{(1+i)} \ .
\label{eq:j}
\end{equation}
Therefore the direct extraction of the $R_{ij}$ from the data is our
starting strategy. As will be seen in the next section, the results are
sensible. The omitted effects, such as correlated efficiencies, etc.
then are attacked as perturbations on this first order analysis, with
PYTHIA/GEANT simulations being the main tool for assessment of their 
importance.

\section{Results}

The charged-particle multiplicity distribution is smooth and is
reasonably fit with a negative-binomial form with a k-parameter in the
neighborhood of 3. The value of $dN/d\eta$ at $\eta=4$, as estimated
from the raw data, is about 3, consistent with a smooth extrapolation
of UA5 data. Errors in these numbers are dominated by systematic
effects, which require more study to determine. Therefore we do not
choose to quote detailed numbers at this time; overall efficiency
determinations and normalizations will be the business of the late, not
early, analysis program.

As described in the previous section, our efforts have been
concentrated on the DCC search via the factorial-moment robust
variables, where the absolute efficiencies play a less central role in
the first-order analysis. The very preliminary values of the low-order
$R_{ij}$'s is given in Table I, together with the expectations from
simulations for generic and/or pure DCC pion production. These results
show no significant dependence upon running conditions, including the
presence or absence of the $x_{F}=0.5$, $0.9$ tags. Detailed limits on
the fraction of DCC allowed by the data must await better understanding
of systematic errors, as well as more detailed modeling of DCC
production mechanisms, but it appears that we are already sensitive to
10--20 percent DCC admixtures.
  
\begin{table}[t]\caption{Values of $r_{ij}$ from the data and Monte 
Carlo\label{tab:data}}
\vspace{0.4cm}
\begin{center}
\begin{tabular}{|l|c|c|c|c|}
\hline 
   & PYTHIA & pure & DCC   &     \\
$r_{ij}$ & and  & DCC   & and &  Data  \\
   & GEANT   &    & GEANT &      \\ \hline \hline
$r_{11}$& $1.01\pm.02$ & $0.500$ & $0.56\pm.01$ & $0.98\pm.01$ \\
$r_{21}$& $1.02\pm.05$ & $0.333$ & $0.40\pm.03$ & $0.99\pm.02$ \\
$r_{31}$& $1.09\pm.14$ & $0.250$ & $0.34\pm.05$ & $1.03\pm.04$ \\ 
\hline\hline
\end{tabular} 
\end{center} 
\end{table}

\section{Outlook}

The basic goals of the MiniMax experiment are being met. Good data on
charged-particle and converted-photon spectra have been acquired, and a
preliminary search for DCC carried out. A search for unusual events
such as Centauro and JACEE has also been made. Nothing singular has been
seen, but firm conclusions depend upon validating our estimates of
detection efficiency, especially for converted photons. We should in
the future be able to contribute to the study of intermittency, and
will attempt to measure the momentum spectra of $K_{S}$'s and $\Lambda$'s.

\section*{Acknowledgment}
This work was supported in part by the Department of Energy, the
National Science Foundation, the Guggenheim Foundation, the Timken
Foundation, the Ohio Supercomputer Center, and the Case Western Reserve
University Provost's Fund.

\end{document}